\documentstyle{mn}

\def\UV{\hbox{$U-V$}}
\def\BR{\hbox{$B-R$}}

\def\VI{\hbox{$V-I$}}
\def\RI{\hbox{$R-I$}}
\def\VR{\hbox{$V-R$}}
\def\gs{\mathrel{\raise0.3ex\hbox{$\scriptstyle >$}\kern-0.70em %
\lower0.71ex\hbox{{$\scriptstyle \sim$}}}}
\def\ls{\mathrel{\raise0.3ex\hbox{$\scriptstyle <$}\kern-0.70em % Less
\lower0.71ex\hbox{{$\scriptstyle \sim$}}}}
\def\et{\hbox{\it et al.}$\,$}

\title[Lensing Studies of Clusters: I -- Faint Galaxy N(z).]
{Gravitational Lensing of Distant Field Galaxies by\\
Rich Clusters: I. --  Faint Galaxy Redshift Distributions.}

\author[Smail et al.]
{Ian Smail,$\!\!$\thanks{Present address: Caltech 105-24, Pasadena CA
91125.} Richard S. Ellis\thanks{Present address:  Institute of Astronomy,
Madingley Road, Cambridge CB3 0HA, UK}  \& Michael J. Fitchett\thanks{Present
address:  Medical School, University of Newcastle, Framlington Place,
Newcastle, UK} \\ Physics Department, University of Durham, South Road, Durham
DH1 3LE, UK
}

\date {Received 1993 --- --; in original form 1993 December 10.}

\begin{document}
\label{firstpage}
\maketitle

\begin{abstract}
{}From deep optical images of three clusters selected by virtue of their X-ray
luminosity and/or optical richness (1455+22; $z=0.26$, 0016+16;
$z=0.55$ and 1603+43; $z=0.89$), we construct statistically-complete
samples of faint field galaxies ($I \leq 25$) suitable for probing the
effects of gravitational lensing. By selecting clusters across a wide
redshift range we separate the effects of the mean redshift
distribution of the faint field population well beyond spectroscopic
limits and the distribution of dark matter in the lensing clusters. A
significant lensing signature is seen in the two lower redshift
clusters whose X-ray properties are well-constrained. Based on these
and dynamical data, it is straightforward to rule out field redshift
distributions for $I \leq 25$ which have a significant low redshift
excess compared to the no evolution prediction, such as would be
expected if the number counts at faint limits were dominated by low-$z$
dwarf systems.  The degree to which we can constrain any high redshift
tail to the no evolution redshift distribution depends on the
distribution of dark matter in the most distant lensing cluster.  In
the second paper in this series, we use the lensing signal to
reconstruct the full two-dimensional mass distribution in the clusters
and, together with high resolution X-ray images, demonstrate that their
structural properties are well-understood. The principal result is
therefore the absence of a dominant low-$z$ dwarf population to $I \leq25$.
\end{abstract}

\begin{keywords}
cosmology: observations -- clusters: galaxies: evolution --
galaxies: formation -- galaxies: photometry -- gravitational lensing.
\end{keywords}

\section{INTRODUCTION}

The surface density of faint galaxies is significantly in excess of
predictions based on extrapolating to large redshift the known local
properties of field galaxies, under the assumption of no evolution
(Kron 1978, Peterson \et 1979, Tyson \& Jarvis 1979). The deepest optical
counts are inconsistent with both modest or no evolution irrespective
of the cosmological geometry (Tyson 1988, Yoshii \& Takahara 1990,
Metcalfe \et 1990, Lilly \et 1991) and reveal a gradual bluing with
increasing apparent magnitude and no convincing turn-over to $B \sim$ 28
(Metcalfe \et 1993).

The high surface density of blue light implies a star formation rate
sufficient to yield a significant fraction of the metals in disk
galaxies today (Cowie 1990). The nature of the population dominating
the counts beyond $B \sim$ 22 is thus of considerable interest and
depends critically on its redshift distribution. To the limits
attainable with high throughput spectrographs on 4-m class telescopes
($B \sim 24$), no significant departure from the predicted no evolution
{\it shape} of the redshift distribution for $B$-selected samples has
yet been seen (Broadhurst \et 1988, Colless \et 1990, 1993, Cowie \et
1991, Glazebrook \et 1993). The most rigorous statement on the redshift
distribution can be made to $B$=22.5 (Colless \et 1993a) where the
incompleteness in a sizeable sample is less than 5\%. At $B \sim 24$
galaxies, where incompleteness remains $\simeq$15\% (Glazebrook \et
1993), a proportion could be in a high redshift tail with $z \geq 1$,
but it is important to note that the incompleteness is negligible
compared to the factor of $\times$4-6 by which the counts exceed the no
evolution prediction. Notwithstanding the incompleteness, the bulk of
the excess population of blue sources, if it exists as a separate
entity, must lie within a volume consistent with the no-evolution
prediction.

Determining the nature of the excess population
is hard because of the difficulty of identifying
representative examples for scrutiny. So long as the excess population
is statistically-defined, physical properties such as luminosity
functions or clustering scale-lengths cannot be reliably determined.
Following Broadhurst \et (1988)'s suggestion that the excess is
associated with galaxies with intense [O II] spectral emission, Cole
\et (1993) found the excess population is co-spatial with its
quiescent counterpart and Colless \et (1993b) find such sources are
often double systems. Such observations are {\it consistent} (but by no
means prove) that star-formation induced by merging may simultaneously
explain the counts and redshift data (c.f.\ Broadhurst \et 1992). On
the other hand, Efstathiou \et (1991) and subsequent workers
(e.g.\ Couch \et 1993), present convincing evidence for a marked
decrease in the angular clustering of $B\simeq26$ galaxies which may
support an alternative viewpoint that the blue light arises in recent
star formation in a separate dwarf galaxy population whose present day
counterparts cannot be found (c.f.\ Babul \& Rees 1992).
Some support for this model has come from limited spectroscopic
surveys which have concluded that there is an excess of dwarf systems
at the required redshifts (Cowie \et 1991, Tresse \et 1993).

The importance of the angular correlation function studies of faint
field samples lies in the fact that the virtually {\it all} of the
sources at $B$=26 represent the excess population i.e.\ difficulties in
identifying the excess populations are largely removed. If it were
possible to determine redshifts and luminosities for such a faint
sample, even if only statistically, significant progress could be made.
If the counts were dominated to the faintest limits by a recent era of
dwarf galaxy formation, as proposed by Babul \& Rees, conceivably the
median redshift would hardly change for samples fainter than
B$\simeq$24.  For a simple merger model, the median redshift closely
tracks the no evolution prediction (c.f.\ Broadhurst \et 1992), whereas
if a significant fraction of $B$=26 galaxies originates in a primordial
population there would be a rapid increase in the median redshift.

Unfortunately, conventional optical spectroscopy is rapidly approaching
a hard faintness limit for two reasons. State of the art faint object
spectrographs such as LDSS-2 (Allington-Smith \et 1993) and MOSIS
(Le F\`evre 1993) secure redshifts to $B$=24 in 4-6 hour exposures.
Even with 10-m class telescopes, it will be painful to push the limits
much beyond $B$=25. More importantly, Glazebrook \et (1993) demonstrate
convincingly how, as [O II] 3727 \AA\ is redshifted beyond 8000 \AA\
(for sources with $z>1$) no useful diagnostic features can be seen
in the optical region resulting in severe redshift incompleteness in
any $z>1$ tail. What is needed, therefore, is an independent
method for determining the mean cosmological distance to a sample
substantially fainter than $B$=24.

In this paper we describe how the gravitational lensing signal produced
by rich clusters at different distances can constrain the redshift
distribution of the faint galaxy population. The relevant samples are
chosen to have $I \leq 25$ (corresponding approximately to $B \ls
27$).  The technique is based on the weak distortion of background
field galaxies first explored in a pioneering paper by Tyson \et
(1990). We have, however, extended the method, not only by imaging the
field population to the same depth through several clusters at
different distances, but also, significantly, by verifying the relative
distribution of dark matter in the lensing clusters using a new
inversion technique developed by Kaiser \& Squires (1992). The latter
result forms the basis of the second paper in this series (Paper II,
Smail \et 1994), which should ideally be read in conjunction with this
paper.

A plan of this paper follows. In Section 2 we briefly review the lensing test
proposed. This serves to explain in more detail the logic of this paper and
its companion article. In Section 3 we discuss the new observations including
target selection, data acquisition and reduction. Section 4 presents various
statistical tests we have applied to the faint catalogues in the context of
model redshift distributions. Our constraints are discussed in Section 5 and
our conclusions are presented in Section 6.

\section {The Lensing Method and Previous Work}

Our method to determine the mean distance to a $I \leq 25$ sample
works as follows. The gravitational lensing of background galaxies
by the cluster potential produces a coherent pattern of image distortions
orthogonal to the cluster radius vector (Grossman \& Narayan 1988).
Although the weak signal is superimposed upon intrinsic ellipticities and
orientations of the population, its coherent nature can be used to overcome
the low signal to noise inherent in the statistics of faint image shapes.
Of course, neither faint cluster members nor sources foreground to the
cluster contribute to the lensing signal.

The most elementary test measures the proportion of objects to a fixed
apparent magnitude limit aligned tangentially to the radius vector to
the lens centre. In the idealised case of a sample of identical
lenses at different redshifts, $z^{i}_{lens}$, i=1,..n, the variation in the
the fraction of aligned images with redshift delineates the shape of the
field galaxy redshift distribution, $N(z)$. In practice, of course,
clusters have a variety of lensing powers and the observed fraction is
controlled not only by the combination of the field $N(z)$ and $z_{lens}$,
but also by the amount and concentration of mass in the lens $M(r)$.
To decouple these two factors, more complex analyses are required.

The first stage of complexity is to allow some freedom in the core
radius, $r_c$ (kpc) and the depth of the gravitational potential well,
parameterised by $\sigma_{cl}$ (km sec$^{-1}$) the velocity dispersion of the
cluster, in a given cluster according to a simple isothermal model.
These parameters can be constrained to some extent from X-ray imaging
data and galaxy dynamics. By applying a joint likelihood
technique across all 3 clusters, each of which has been imaged in the
same conditions, we can test whether the lensing signals are consistent
with a sequence of `test' field redshift distributions, $N(z)$.

It may be, however, that the distribution of dark matter in a cluster
bears little relation to that observed for the X-ray gas and cluster
members. In such a situation, the test describe above would give
misleading results. The companion paper to this article shows how
the lensing signal measured {\it across} the cluster can be inverted,
using a new technique developed by Kaiser \& Squires (1992), to define
a projected 2-D map of the lensing mass at a resolution adequate to check
its concentration. Although the technique does not yield an
{\it absolute} estimate of the total cluster mass, the results show how
closely the dark and baryonic matter are distributed. This important
result is sufficient to remove the ambiguities essential to determining an
accurate estimate of the median redshift for a $I \leq 25$ field sample.

Our method is qualitatively different to previous lensing probes of the
faint galaxy distribution, and thus we briefly review those studies in
the context of our paper. Tyson \et (1990) presented a pioneering
analysis of two X-ray luminous clusters: Abell~1689 ($z=0.18$, L$_X =
1.7$ 10$^{45}$ ergs sec$^{-1}$, $\sigma_{cl} =$ 1800$\pm$200 km
sec$^{-1}$) and Cl1409+52 ($z=0.46$, L$_X = 9.2$ 10$^{44}$ ergs
sec$^{-1}$, $\sigma_{cl} \sim$ 3000 km sec$^{-1}$). The small CCD
format available at the time restricted imaging to radii $r \leq$ 380
and 500 kpc respectively (we adopt $H_o$=50 km sec$^{-1}$ Mpc$^{-1}$,
$q_o$=0.5 throughout). Samples were selected with $B \in [22,26]$
(equivalent to $I \in [20,24]$) but only the bluer galaxies showed
alignment tangentially to the cluster centres.  However, the excess
aligned component is relatively small, amounting to $\simeq$30-40
galaxies in A1689 and only 12 in Cl1409+52. Using this signal, Tyson
\et claimed that at least 70\% of the $B \in [22,26]$ population has $z
\geq 0.9$. Applying a statistic based on the alignment signal they also
derived radial `mass' profiles for the clusters but it transpires these
profiles represent the surface potential (c.f.\  Kaiser \& Squires 1992).

Our study extends the analysis of Tyson \et (1990) by constructing
a more extensive sample of faint galaxies with large format CCD detectors
and applying a variety of new analytic techniques. There are also some
important strategic differences. Firstly, by selecting in $B$ and restricting
the cluster redshifts to $z \leq 0.5$, Tyson \et would not be sensitive
to a genuinely high redshift population whose Lyman limit would have
shifted beyond the observing passband. By selecting in a near-infrared
passband and using long baseline colours, we can rectify this and
more easily remove contaminating cluster members.  The second important
difference is increased depth. By probing a magnitude deeper in good
seeing, we reach a surface density of sources sufficiently high
to allow recovery of possible substructure in the cluster mass distribution
using the Kaiser \& Squires inversion technique. Simulations show this
would not be possible at Tyson {\it et al.}'s brighter limit.

Very recently, Kneib \et (1993) have utilised a rather different method
to constrain the redshift distribution of $B\simeq27$ galaxies. They
select a single cluster, Abell 370 ($z$=0.37), whose mass profile is
well-constrained from arcs of known redshift (Soucail \et 1988).  They
visually identify a candidate list of likely arclets and attempt to
`invert' the lens equations to derive their cosmic distances {\it
individually}.  Whilst a promising technique, it is important to
recognise that this approach cannot explore all regions of redshift
space with uniform sensitivity. Thus whilst the individual arclet
redshifts may be correct, by using a single cluster an unbiased $N(z)$
appropriate for a $B \leq 27$ sample cannot be simply constructed.

\section {The Data}

\subsection{Observational Considerations}

The lensing signal we seek is intrinsically very weak and could  easily be
affected by systematic errors. To account for such errors, we have simulated
images of clusters taking into account all likely observational effects. These
simulations, discussed below, are used to calibrate the statistics we apply to
the real datasets. Foremost is the need to measure ellipticities of faint
galaxies over a wide field. The typical scale length of a $I\simeq25$ galaxy is
0.3-0.7 arcsec, thus sub-arcsecond seeing in the selection passband is
critical. In addition, the pixel scale must sample the seeing disk
appropriately to eliminate pixellation effects in the ellipticity measurements
(pixels $\ls$ 0.3 arcsec). With a large format EEV CCD, the f/4 TAURUSII
imaging Fabry-Perot system on the 4.2-m William Herschel Telescope (WHT)
has 0.27 arcsec sampling of a 5 $\times$ 5 arcmin field, making it ideally
suited to this project.

The two passbands ($V$ and $I$) were selected to provide a
colour baseline consistent with the sensitivity of the available large format
EEV CCD. To reach a surface density of 40 arcmin$^{-2}$ in the $I$ band
requires a completeness limit of $I$=25 (Lilly \et 1991). Data in
$V$ permits the discrimination of cluster members as well as exploration of
a possible variation of the lensing signal with colour. Typically,
\VI $\sim$ 1.5 leading to a $V$ completeness limit of $V$=26.5.

Excellent seeing is only needed for the $I$ detections which are then
used to provide statistical estimates of shapes and orientations on
a galaxy-by-galaxy basis. Within the galaxy image, we estimated a
minimum signal to noise per pixel of $\simeq$2.5 corresponding to
50 sigma over a 1.5 arcsec FWHM. To achieve this signal to noise at the
chosen completeness limit required on-source integrations of 20 ksec in
$I$ and 10 ksec in $V$ per cluster.

Most published moderate and high-$z$ cluster catalogues
were identified either from
peaks in the projected surface density of optical galaxies (Abell \et 1989,
Gunn \et 1989, Couch \et 1991), or from deep X-ray observations (Henry \et
1992).  Although the negative evolution claimed for the number density of
luminous X-ray clusters at relatively low redshifts (Edge {\it et al.\ }1990)
seems in conflict with the abundance of high redshift optical clusters (Couch
\et 1991), we conclude X-ray observations still provide us with a tracer of
the most massive systems at a given epoch and should be less sensitive to
projection effects. Accordingly, we chose 3 clusters (1455+22, $z=0.26$;
0016+16, $z=0.55$ and 1603+43, $z$=0.89) foremost on the basis of their
X-ray luminosities.
Further details of the clusters are provided in later sections
(see also Table~1).

Observations were made in two runs in July 1990 and May 1991 on the
WHT; the journal is presented in Table~2. During the first run we encountered
photometric conditions with very good seeing. However, during the second
portion of the second run the seeing deteriorated beyond the limit
we considered suitable for this study (1.0 arcsec FWHM). Accordingly
the data taken
during this period was only used for photometric work and was not used
to analyse images for the lensing analysis.

The observations comprised multiple $\ls$ 1000 sec exposures of two
target clusters in a single passband each night. The exposures were
`dithered' on a rectangular grid with 10-15 arcsecond spacing and, by
combining data for both clusters, a sky flatfield for the entire night
was constructed and used to process the images. Numerous shorter
exposures of standard stars were also taken at regular intervals through
out the night to track variations in the transparency. Further twilight
flatfields were taken each night. The transparency was very stable
with photometric zero point errors on the final frames below 0.01 mag.

The reduction of in-field dithered images is a more complex affair
than normal image reduction, especially for a rich cluster containing many
bright large galaxies. The reduction was performed using IRAF and consisted
of the following steps.  1) After bias subtraction and trimming,
large scale gradients across the images were removed using the twilight flats.
2) An object detection algorithm (FOCAS, Jarvis \& Tyson 1981) was then
used to replace all pixels contained within bright objects with random sky
values drawn from surrounding regions. 3) The cleaned frames were then
median-combined to give `superflats' which contain both the pixel-to-pixel
sensitivity variations and residual large scale variations arising from the
mismatch of the twilight flatfield and the actual flatfield.  5) The
pre-cleaned
images were then flatfielded using the superflats to produce the final
reduced images.

A comparison of offset frames of the same field after alignment
revealed an uncorrected radial distortion in the positions of all
sources. Investigations revealed this to be an inherent feature of the
focal reducing optics. Uncorrected this would produce a serious
reduction in the off-axis image quality. To remove this distortion,
every image was geometrically remapped according to a system of
fiducial references defined by a large number ($\gs$100) of objects
distributed over the frame.  A spline surface was fit to the X and Y
distortion vectors and each pixel was remapped according to linear
interpolation within this surface.  The remapping successfully removed
the distortion to a level below 0.07 arcsec over the entire frame.  The
remaining residuals appear random and thus should
add incoherently when the frames
are stacked resulting in an uncorrelated rms error in the ellipticity
of a typical faint object of $\leq$0.5\%.

The processed frames were then combined using a broad median algorithm
with an additive scaling offset to account for variations in sky brightness.
This provided two final frames, one in each passband, for each of the three
clusters. The typical 1$\sigma$ surface brightness limit in the two bands are:
$\mu_V = 28.9$ mag arcsec$^{-2}$ and $\mu_I = 27.8$ mag arcsec$^{-2}$.
We now discuss the creation of reliable objects catalogues from these deep
images.

\subsection{Object Selection}

The size and depth of the frames obtained are such that, when
optimizing the analysis technique, it is most efficient to run tests on
a small, but representative, 1$\times$1 arcmin regions (Figure~1). The
FOCAS object detection algorithm is described in detail by Jarvis \&
Tyson (1981) (see also Valdes 1982). The two main parameters which
control the detection characteristics of the algorithm are the
threshold cut (in units of the global sky sigma; $\sigma_{sky}$) and the
minimum object area (in pixels). These parameters were optimised by
visually checking the success rate of faint object detections against
obvious spurious sources. The optimum combination was a threshold of
2.5 $\sigma_{sky}$ per pixel over an area of 10 pixels which corresponds
roughly to a 25 $\sigma_{sky}$ detection within the seeing disk.

Object detection is performed on coadded $V+I$ images, combined such that in
the final image a flat spectrum source ($f_{\nu}\simeq$const -- representative
of the faintest objects) has equal flux contributions from each filter.
Tests show that this approach provides both a fainter detection limit for
images and improves detection of objects with very extreme colours. These
frames are reproduced in Figures 2. After initial detection on the $V+I$
image, the object areas were evaluated and analyzed on the individual $V$ and
$I$ frames. These catalogues were merged to define a final list of sources
detected on both frames.  Any objects whose isophotes touch the frame
boundaries and thus have ill-defined shapes, are rejected at this point.

To determine the effective completeness limit for object detection  in
the individual bands we create a high signal to noise faint galaxy by
median combining a large number of faint galaxy images from the data.
This is then scaled and repeatedly added into a region of the cluster
frame, the detection process is then rerun and the success of detection
of the images as a function of magnitude gives the completeness limit
for the catalogue for that band.  As noted earlier we actually select
from the combined $V+I$ image which will mean that the limits from the
completeness simulations are in fact lower bounds on the actual
completeness of our catalogues.  We adopt a fixed $I$ magnitude limit
in all three catalogues so that a well-defined redshift distribution
can be compared across all clusters.

Standard aperture photometry in a 3 arcsec aperture is then performed
on all the objects using seeing-matched images and aperture colours
calculated. The resulting parameters for each object are its position,
intensity-weighted second moments calculated from the better seeing $I$
image, isophotal $I$ magnitude and \VI\ aperture colour. The colour
magnitude data for each cluster is shown in Figures 3. The 80\% and
50\% completeness limits for the individual catalogues are marked on these
figures. We now discuss the individual clusters and their associated
faint object catalogues.

\subsection{The Lensing Clusters}

\subsubsection{1455+22 ($z$=0.26)}

This cluster (Figure 2(a)) was discovered as a serendipitous source in
the EINSTEIN Medium Sensistivity Survey (Henry \et 1992). On the basis
of broad-band galaxy photometry, it was initially suspected to be a
$z=0.7$ cluster (Schild \et 1980), but subsequent spectroscopy
confirmed a lower redshift (Mason \et 1981). Unfortunately, redshifts
are only available for four members, including the dominant central
galaxy ($z=0.258$). Although the formal velocity dispersion is only
$\sim$700 km sec$^{-1}$, with only four velocities it would be
possible for the observations to be consistent with an intrinsic
dispersion of 1500 km sec$^{-1}$ 40\% of the time. In contrast, it is
one of the most X-ray luminous cluster known: $L_x \sim$ 1.598
10$^{45}$ ergs sec$^{-1}$ in the 0.3--3.5$\,$keV band. The target has
been imaged in deep pointed observations with ROSAT High Resolution
Imager  in parallel with our gravitational lensing study and this image
is presented in the companion paper.

Examining the colour-magnitude diagram for this field (Figure~3(a)) it is
straightforward to locate a well-defined
cluster colour-magnitude relation, which extends over 6 magnitudes to $I$=22.0,
at the appropriate colour for a population of early-type members at
this redshift.  (It should be noted that all 5 objects originally
thought by Schild \et to be at $z=0.7$ lie on this relation). By
comparing object densities on the colour-magnitude plane with those,
suitably scaled, for the other clusters, no statistical excess
representing the cluster was found either beyond $I$=22.0 or bluer than
the colour-magnitude relation. The tightness of the relation over this
range ($\Delta(V-I) = 0.04$ mag) verifies the excellent photometric precision
achieved. Using the test discussed in $\S$3.2, the 80\% photometric
completeness limits were found to be $I=25.3$ and $V=26.5$. An $I=25$
galaxy corresponds to a 17 sigma detection within the seeing disk.

We define our `field' sample to be those sources whose colours lie off
the narrow colour-magnitude sequence of the cluster (marked as open
circles on Figures 3).  This is a fair approximation given the limits on
a blue excess from the object densities on the colour-magnitude plane
and the observation than even the strongest `Butcher-Oemler' clusters
the fraction of cluster galaxies lying outside this narrow
colour-magnitude sequence is typically less than 30\% (c.f.\ Oemler
1992). Two simple checks can be made of this procedure: in Figure 4 we
compare the number counts derived with those of genuine `blank fields'
(Lilly \et 1991) finding good agreement; we also examined the radial
distribution of our `field' objects which reveals no centrally
clustered component.

The above procedure yields 180 early-type members brighter than $I=22$
over the 5$\times$5 arcmin field. Only 17 galaxies in the inner 500 kpc
lie within the range [$m_3,m_3+2$], compared to 48 for the Coma cluster
(Metcalfe 1983). Clearly, 1455+22 is only a third as optically rich as Coma, a
result which is consistent with the poorly-determined velocity
dispersion.  On the other hand, the extremely high X-ray luminosity and
the presence of a large cD galaxy both point to a deep and centrally
concentrated cluster potential.

\subsubsection{0016+16 ($z$=0.55)}

This cluster was discovered by Richard Kron on a 4-m Mayall prime focus IIIa-F
plate (Spinrad 1980). The redshift is $z = 0.545$ (Dressler \& Gunn 1992) and
the rest-frame velocity dispersion derived from 30 members is $\sigma_{cl} =
1324$
km sec$^{-1}$. The cluster has been the subject of several photometric studies
because of Koo's (1981) original claim that, despite the high redshift, there
is little or no population of associated blue members. The large population of
intrinsically red galaxies has been studied by Ellis \et (1985) and
Arag\'on-Salamanca \et (1993). The cluster was the subject of one of the
deepest EINSTEIN High Resolution Imager exposures (White \et 1981) and
has been recently imaged with the PSPC on-board ROSAT.

The cluster has no dominant central galaxy. The core contains three bright
galaxies in a linear structure (Figure~2(b)) and the peak in the galaxy
surface density lies slightly to the south-west of this structure. The
central members define an elliptical form (axial ratio $\sim 0.6$) and
a core radius of $\simeq$330 kpc. The optical counts indicate a richness of
twice Coma, but Ellis \et~indicate this may be an over-estimate because of
contamination by a foreground system at $z\sim0.3$.

The cluster was detected in the deep EINSTEIN HRI exposure with $L_X
\simeq 1.43$ 10$^{45}$ergs sec$^{-1}$ in the 0.5--4.5$\,$keV band. The
best fitting isothermal $\beta$-model has an X-ray core radius of 220
kpc (White \et 1981), although a strong cooling flow would affect this result.
Although the published X-ray map has only a 30 arcsec (220 kpc)
resolution, it too shows a roughly elliptical structure around the
optical centre orientated similarly. Finally, the cluster has a
detectable Sunyaev-Zel'dovich decrement (Birkinshaw \et 1992). In
summary, therefore, the cluster appears to be very rich and centrally
concentrated and ideally suited for lensing studies.

Following the methods discussed for 1455+22, the 80\% completeness limits in
this case are $V$=26.4 and $I$=25.7, and the $I$=25 limit corresponds
to a 19$\sigma$ detection in the seeing disk. Using the colour-magnitude
sequence, the cluster can be detected down to $I$=23.5. Additional colour
information is available from a deep 6 ksec $R$ band service exposure taken
with a large format EEV CCD at the 2.5m INT prime focus. This is adequate to
provide colours to $I$=24 to better than 0.2 mag. It was therefore possible
to double check the colours of cluster members on a (\VR)--(\RI) plane.
The final cluster sample contains 174 galaxies to $I$=23.5 of which 83 lie
in the interval [$m_3,m_3+2$]. Significantly, the colour-selected members
delineate a complex 2-D structure. The main feature is a partial annulus
consisting of four separate clumps surrounding the cluster centre on the west.

The colour distributions shown in Figure~5 demonstrate that whilst we have
successfully removed the bulk of the cluster members, the foreground groups
identified by Ellis \et (1985) are apparent as an excess of objects brighter
than $I$=22 in the $I$ counts. Removal of this  source of
further contamination is not possible with the existing datasets.

Again, the scatter about the cluster colour-magnitude relation is surprisingly
small ($\Delta(V-I) = 0.06$ mag). Since, at $z$=0.55, \VI\ is equivalent to
the restframe \UV\ -- we can directly compare this value with the intrinsic
dispersion seen in Coma ($\Delta(U-V) \leq 0.04$mag) by Bower \et (1992).  At a
lookback-time of 6 Gyr, there appears to be no evidence for a large increase
in this dispersion. If 0016+16 is a representative cluster then within the
framework developed by Bower \et  this yields a lower limit on the epoch of
formation of cluster ellipticals of $z_{for} \gs 3$ (Ellis 1993).

\subsubsection{1603+43 ($z$=0.89)}

A high redshift ($z\simeq1$) cluster was considered essential in
our survey in order to test the possibility of a truly high redshift
($z \geq$ 2) component in the faint counts. Selecting such a cluster
presented little difficulty since very few are known. At the time of its
discovery, 1603+43 was the most distant optically-selected cluster known.
It was discovered by Gunn \et (1989) and the subsequent spectroscopic
follow-up by Dressler \& Gunn yielded redshifts for $\sim$5 cluster members
with a mean redshift of $z=0.895$. The cluster was also included in the
study of high redshift cluster populations by Arag\'on-Salamanca \et (1993).

The very long exposure times on the WHT for this target produce
faint 80\% completeness limits of $I$=25.9 and $V$=26.3 (Figure 2(c)).
The $I$=25 detection limit corresponds to a 21 $\sigma$ detection within the
seeing disk. The colour-magnitude diagram (Figure~3(c)) shows that
the cluster colour-magnitude relation is broader
than in the lower redshift clusters and offset in colour compared to
non-evolving ellipticals at $z$=0.89 (c.f.\ Arag\'on-Salamanca \et 1993).
An excess of objects can be identified to at least $I$=23.5.  To
remove cluster members, we adopted a very broad colour criterion combined
with a faint magnitude cutoff of $I$=24.0. In this way, we identified
70 cluster galaxies, with 33 in the range [$m_3,m_3+2$] i.e.\ a
richness comparable to Coma. The cluster centre
is associated with a prominent `V' of galaxies and the spatial distribution of
members shows a bimodal structure with one peak over the `V' of galaxies in a
frame centre and the second peak lying to the west on the frame border.

This cluster was one of 4 high-redshift targets from the Gunn \et
sample imaged with the ROSAT PSPC (Castander \et 1993). The cluster was
detected within a $2\times2$ arcmin aperture at a 6.5$\sigma$
significance level in a total exposure time of 28 ksec, corresponding
to $L_X \simeq$1.1$\pm$0.2 10$^{44}$ ergs sec$^{-1}$.  Given the
apparent optical richness of this cluster from the $K$-band imaging of
Arag\'on-Salamanca \et , the low X-ray luminosity is somewhat surprising.
Castander \et conjecture that the negative evolution in cluster X-ray
properties claimed at low redshift (Edge \et 1990) continues in more
distant samples. Whilst these observations give no indication of the
likely concentration of the mass in 1603+43, they do show that it has a
much higher X-ray luminosity than many of the the lower redshift Couch \et
sample -- some of which have gravitational arc candidates (Smail \et 1991).

The optical richness of this cluster is consistent with the high
spectroscopic identification rate for members and, when combined with
the X-ray luminosity, provides  good evidence that the cluster is
massive. In the absence of the strong evolution observed the low
redshift $\sigma_{cl}$--L$_X$ relation would yield a rest-frame 1-D
velocity dispersion of $\sigma_{cl} \gs 800$ km sec$^{-1}$.  If the
observed evolution arises from effects other than growth of the cluster
potential wells, as has been proposed by Kaiser (1991), then this value
is a lower limit to the dark matter's velocity dispersion (the relevant
quantity for the lensing studies).

\subsection{Field Colour-Magnitude Distributions}

Figure~4 shows the field colour-magnitude distributions constructed by the
procedures discussed earlier for each of the 3 cluster areas. All
show the well-known trend to bluer colours at fainter magnitudes.
The median colour for the entire field sample brighter than $I=25$ is
\VI $= 1.55\pm0.10$. The lower envelope to the colour distribution is
\VI $\sim$ 0.9, similar to that of a flat spectrum source. Interestingly,
the number of objects with flat spectrum colours increases rapidly
beyond $I \sim 23$ ($B \sim 25$). Previous workers (Tyson 1988, Cowie
\et 1989) claimed a discontinuity in the photometric data at about
this point. Certainly, the colour distributions brighter and fainter
than $I=23$ are highly inconsistent with being drawn from the overall parent
population. However, when allowance is made for the bluing of the entire
population, the shapes of the two distributions are very similar.

The deepest uniform $I$ sample  consistent with the detection limits
across the 3 clusters is  $I_{iso}=25.0$ giving $>$95\% completeness
limits in all 3 clusters.  This limit corresponds
to a minimum detection significance in the $I$ band of $\sim$17$\sigma$
in the seeing disk. when combined with a similar detection requirement
in the $V$ frame this creates a very robust sample with which to work.

\subsection{Estimating Image Parameters of Faint Field Galaxies}

Figure~1 illustrates a random 1$\times$1 arcmin test area taken from the
1455+22 field. The frame contains 40 objects brighter than $I_{iso}=25$
in the field and those with $I_{iso} \in 24-25$ are marked. The lensing
technique relies upon our ability to estimate the ellipticities of these
faint objects.

The problem of measuring reliable ellipticities for faint objects remains an
area of active research. Intensity-weighted second moments (as used in FOCAS)
can yield reliable ellipticity and orientation estimates for bright sources
but, for the faintest objects under consideration here, the outer isophotes are
heavily influenced by noise. For this reason, the intensity-weighted and
unweighted moments give very similar results.

To circumvent this, we developed an alternative approach. Instead of
using the detection isophote to define pixel membership for an object,
we select a circular aperture and, to reduce the noise from the outer
regions, a {\it radial} weighting function is applied when calculating
the second moments in this aperture. The optimal weighting function for
a particular object then has the same profile as the object. To
simplify matters, we adopted a generic circular Gaussian with a
variable width as a weighting function -- this simplification has been
shown to be reasonable (Bernstein, priv.\ comm.). The width was
determined from the intensity-weighted radius of the object broadened
by convolving with the seeing psf. We refer to moments measured using
this algorithm as `optimally weighted'.

Two separate tests were undertaken to estimate the reliability of the
ellipticity measurements for the faintest objects in our sample. The
first test involved estimating the ellipticity errors of $I=25$
galaxies using simulations. The second test measured the scatter in an
individual measurement from two independent observations of the same
field. The simulations consisted of a large number of artificial frames
populated by objects with known ellipticities.  For the comparison
test, individual exposures comprising the final 1455+22 $I$ frame were
combined to create two independent frames each with a total exposure
time of 9.5 ksec. These were then analysed and the resulting catalogues
matched to allow comparison of the measured image parameters
(Figure~6). Both tests have some drawbacks -- the simulation results
are dependent upon the form of the galaxy profile used, while the real
observations are of necessity shallower than the final image.

In the simulations, the intensity-weighted FOCAS moments provide an
unbiased and reasonably accurate estimate of the input object
ellipticity ($<\!\!  \Delta\epsilon \!\!> = 0.16$). The optimally-weighted
moments are systematically rounder by about 0.1 than the input
(Figure~6).  However, the comparison test showed that the
optimally weighted moments have a roughly four-fold reduction in the
scatter in the ellipticities measured for an object from both frames
($<\!\!  \Delta\epsilon \!\!> = 0.04$ versus $<\!\!  \Delta\epsilon
\!\!> = 0.16$ for the objects with $I_{iso} \in 24-25$).  A similar
reduction in scatter and introduction of a systematic offset has also
been reported by Bernstein (priv.\ comm.).

In our analysis we use the more efficient optimally-weighted moments
for tests where the systematic bias introduced could be
modelled (such as the mass mapping presented in Paper II),
otherwise the intensity weighted FOCAS moments were used.

\section{Statistical Analyses of Gravitational Lensing}

\subsection{Model Redshift Distributions}

Our primary goal is to use gravitational lensing as a tool to
constrain the redshift distribution of faint field galaxies well
beyond the spectroscopic limits of the largest current telescopes.
As we are mainly concerned with establishing a {\it statistical} result
for the mean distance to the faint population at $I \leq 25$, we have
tested our lensing signals against three model redshift distributions, $N(z)$,
which encompass the various physical models discussed in $\S$1.

The three model distributions adopted for the $I\leq25$ samples
are: (i) the {\it no evolution} (`N.E.') prediction which maintains
a reasonable fit to the deepest spectroscopic observations thus far
and might be considered an appropriate model for the merger-induced
star formation picture (Broadhurst \et 1992); (ii) the {\it shallow}
prediction which maintains the form of the distribution observed at
$I$=21 irrespective of the limiting magnitude. Fainter than $I$=21,
galaxies simply pile up in the same redshift range as might be
expected if there was a well-defined era of recent dwarf formation
(Babul \& Rees 1992). (iii) Finally we have a {\it deep} prediction which
includes a significant proportion of galaxies with $z>1$ as originally
claimed by Tyson \et (1990). We adopted the distributions of White
\& Frenk (1991) which are based on a hierarchical model for galaxy
formation and transformed roughly from $B$ to $I$ using a fixed colour
term.

The 3 model redshift distributions are summarised in Figure 7.  The
potential of our clusters to distinguish between these models can be
examined by considering the proportion of $I\leq25$ galaxies lying
beyond our clusters. For the 3 model $N(z)$ the fractions behind
1455+22, 0016+16 and 1603+43 are, respectively, Shallow (63\%, 1\%,
0\%), NE (96\%, 69\%, 20\%) and Deep (97\%, 83\%, 65\%). Whilst a
continuum of intermediate possibilities are physically plausible,
particularly between the no evolution and deep cases, the 3 models are,
we believe, sufficient for this exploratory study.

\subsection{The Lensing Tests}

While image parameters have been determined for statistically-complete
catalogues of field galaxies in the 3 cluster areas, we still have
to develop algorithms for estimating the coherent lensing signal.
In recent years a number of statistical methods have been developed to
analyse the weak lensing of faint galaxies by rich clusters (Kochanek 1990,
Miralda-Escud\'e 1991a, 1991b, Kaiser \& Squires 1992). In general, these
methods aim to derive the mass profile of the lensing cluster, rather than
the properties of the faint galaxy population -- which are assumed to be
known.

The analyses fall into two main classes: parametric likelihood tests which
assume some functional form for the relative mass distribution in the lens
and then attempt to determine the most likely values of the model
parameters (Kochanek 1990, Miralda-Escud\'e 1991a, 1991b) and non-parametric
tests which  directly infer the 2-D projected mass distribution
(Kaiser \& Squires 1992). The former methods are capable of testing the faint
galaxy properties, whereas the latter methods are
better suited for investigating the {\it relative} distribution of mass in the
lensing cluster.

Throughout this paper, we will assume our lenses can be modelled
by a spherically-symmetric non-singular isothermal sphere parameterized by a
core radius, $r_c$, and a rest-frame one dimensional velocity dispersion,
$\sigma_{cl}$.  We chose this simplification initially to make progress in the
absence of any other information. As we explained in $\S$1, however, the
companion article (Paper II) presents the non-parametric analyses using the
Kaiser \& Squires statistic and those results allow us to test directly
the parametric methods adopted in this paper. The uncertainties in assuming
the clusters can be parameterised by simple isothermal models are reviewed
in that paper.

\subsection{Parametric Methods}

The parametric tests compare the observed distributions of image
parameters with those calculated for a family of lensing clusters
for each of the various $N(z)$. These `model' distributions were first
calculated using an {\it analytic} prescription of the lensing effect of a
given cluster, assuming the data has very simple noise properties.
However, in tests, we found that although this method is sensitive to
the redshift distributions, it yields cluster parameters which are
systematically offset from their true values.
This is presumably because the real data contains systematic effects
not represented in the analytical treatment. To correct for this degradation,
we undertook more realistic {\it simulations} which attempt to include all the
likely sources of observational noise and estimate, as accurately as possible,
to calibrate the offset in the cluster parameters.

To determine the suitability of a given $N(z)$ for a given set of
cluster parameters, we adopted a simple maximum likelihood technique.
Consider two redshift distributions, a test hypothesis (say, the deep
case) and a null hypothesis (the no evolution case). For both we
estimate the probability that the observed dataset for a given cluster
can be reproduced
according to a family of lens models. Applying the maximum likelihood
method to each hypothesis will yield two estimates of $\sigma_{cl}$,
denoted $\hat \sigma_0$ and $\hat \sigma_1$, two estimates of $r_c$
($\hat r_{c0}$ and $\hat r_{c1}$) and two probabilities $\hat p_0$,
$\hat p_1$. These probabilities are determined by comparing the
observed image orientations, ellipticities and radial positions from
the lens centre with those predicted by the models ({\it c.f.} Smail
\et 1991). We compare the hypotheses by constructing the likelihood
ratio $\Lambda=\hat p_0/\hat p_1$. If the ratio is large ($\Lambda \gg
1$), the alternate hypothesis is rejected in favour of the null
hypothesis.  This approach can obviously be extended to test the
relative likelihoods for our three model $N(z)$.  In addition to
selecting the most likely $N(z)$, the maximum likelihood method returns
preferred values for the lens parameters for each cluster.

\subsubsection{Analytical Solutions}

The analytical test works as follows. For each $N(z)$, a combination
of core radius and velocity dispersion for the lens are chosen
from a grid of values.
Galaxies are then drawn randomly from the hypothesised redshift
distribution. The model galaxies are distributed uniformly across the
source plane with ellipticities drawn from the observed blank field
distribution and random orientations. The image distortion arising from
the lens is calculated using the formalism of Miralda-Escud\'e (1991a)
which yields the radial position ($r$), the orientation relative to the
lens centre ($\theta$), and the ellipticity ($\epsilon$) of the model
image. The procedure is repeated until there are sufficient objects to
allow a fair comparison of the model distributions with the
observations.  A linear Kolmogorov-Smirnov test compares the observed
and predicted distributions and the final likelihood that the model
could create the observations is determined by combining the
probabilities ($\log \hat p = \log (p_{\epsilon} p_{\theta} p_{r})$).
The test is extremely powerful when applied to strong lensing systems.
However, as the lensing signal diminishes so does its distinguishing power.

Figure 8 shows how accurately our lensing test can determine
the correct input cluster parameters for two different kinds of
simulations. Both simulations adopt the observational parameters
(cluster redshift, frame size and field galaxy magnitude limit) for the
1455+22 dataset and assume, as input, that the cluster is a spherical
system with $\sigma_{cl}$=1400 km sec$^{-1}$ and $r_c$= 100 kpc
and the galaxies are drawn from the no evolution redshift distribution. The
logarithmically-spaced probability contours show the derived cluster
parameters assuming the observed galaxies are drawn from either the no
evolution ($H_O$) or shallow ($H_1$) redshift distributions. The filled
circle denotes the correct input value in the $\sigma-r_c$ plane. In
the top panels, the catalogue was constructed using the analytical
formalism of Miralda-Escud\'e (described above), whereas in the bottom
panels the simulations attempt to allow for as many of the
observational selection effects as possible by constructing a realistic
{\it frame} of the simulated cluster (see $\S$4.3.2 for a detailed
description).

In the case of the analytical models, the test readily returns the
correct redshift distribution: the probability ratio is $\log \hat p_1
/ \hat p_0 \equiv \log \Lambda  < -10$.  In addition the input lens
parameters are correctly recovered. The shape of the likelihood
contours can be understood in terms of a trade-off between an increase
in $\sigma_{cl}$ -- which strengthens the lens -- against an increase
in $r_c$ -- which weakens it. The shifts between the contours for the
two redshift distributions arises because more distant galaxies are
more easily distorted. Examining the individual distributions, we find
greatest power comes from the orientations which constrain the solution
to lie somewhere along a slanted locus. The ellipticities and radial
positions then confine the solution to a point on this locus. Whilst
the overall likelihood is derived assuming that the probabilities from
the three K-S tests for $r$, $\theta$ and $\epsilon$ are independent,
this is not a critical assumption given the dominant power of the
orientation distribution.

For the catalogue from the simulated frame, the test still correctly
distinguishes between the two possible redshift distributions with $\log
\Lambda  = -1.8$. However, a systematic offset in the best fit lens parameters
appears. The calibration and correction for this offset is the
motivation for creating simulated frames and its source is
discussed below.  The ability of the test to determine the correct lens
parameters is very sensitive to the strength of the observed lensing signal.
For weak signals the likelihood peak flattens and while the test can still
determine the correct redshift distribution, the lens parameters become
less meaningful.

The analytical catalogue (created in the same manner as the
analytic models) obviously disregard a number of complications. The
effects of noise on the image measurements and the degradation of the
induced distortion by seeing are ignored. Both these effects will
introduce a systematic error in the measured ellipticity.  However,
their effect on the image orientations will be random. By concentrating
on the image orientation and radial distributions it is hoped that the
effects of these systematic errors will be minimized.  Furthermore, the
fixed source magnitude limit results in a paucity of
objects in the lens centre on the image plane. This is because
amplification bias has been neglected, this acts to populate this
region by magnifying galaxies fainter than the observation's magnitude
limit into the sample.  The combined effect of all these processes is
to reduce the observed lensing signal in the simulated frames (Figure~8).
By using simulated observations this degradation can be calibrated and
the observations corrected for it.

\subsubsection{Simulated Frames}

To calibrate the statistical tests applied to the real observations we
simulate a set of frames which are analysed in the same manner as the
real data using FOCAS. This approach was chosen to cater for most of the biases
that are likely to occur in the data which cannot be handled
analytically. Each of these may degrade the strength of the lensing
signal causing systematic errors in the derived lens parameters. The
most obvious effect is atmospheric seeing, but sky noise and undetected
merged images also contribute. Underlying correlations in the data due
to redshift dependency of certain image characteristics are also a
concern unless properly modelled. The lensing is performed using the
same technique as in the analytic simulations with a number of
additional features:

\noindent{$\bullet$}  The effects of seeing, pixellation and sky noise
on the measurements are included in the simulated image to match
those appropriate for a particular observation.

\noindent{$\bullet$} Instead of a cumulative $N(z)$ to a given magnitude
limit, a differential distribution is used which allows empirical
control of changes to the form of the $N(z)$ as a function of magnitude
allowing us to easily model the effects of amplification bias in our
observations.  Furthermore, as the distortion at a particular radius on
the source plane increases with redshift more distant sources will tend
to be more distorted.  However, as the signal to noise in a given
ellipticity measurement decreases for fainter objects it is important
that the faintest, possibly most distant and hence most strongly lensed
objects are most effectively degraded.   Ideally, we would also like to
include any correlation of source size or ellipticity with redshift but
there is currently no observational data on either of these
correlations. Images to $I$=23 unaffected by atmospheric seeing were
kindly supplied by the HST Medium Deep Survey (MDS) Team and scale
lengths for sources selected randomly from their first deep image
(Griffiths \et 1992) measured for this purpose.

\noindent{$\bullet$} The effects of crowding on image detection and the
distortion of image isophotes by undetected faint images are also be
included. The latter effect is incorporated by lensing galaxies fainter
than the adopted magnitude limit (to $I$=27), their number being
determined from extrapolating the observed counts. Detected objects
fainter than $I$=25 are then discarded after the FOCAS analysis. The
sources are uniformly distributed on the source plane and the inclusion
of the fainter sources thus allows modelling of amplification bias.
Current data implies a very weak two point correlation function for
such faint objects (Efstathiou \et 1991, Couch \et 1993) and so the
crowding effects will not be appreciably underestimated.

Of course, again, there are shortcomings with these simulations.
The most obvious is that the technique used to lens the galaxies does
not produce any curvature in the final images.  This is unimportant for
the majority of images, but rare strongly-lensed images (giant arcs)
close to the cluster centre are not well modelled.  None of the
three datasets contains such a giant arc and so an upper ellipticity
cut-off can be applied to both the simulations and real data to remove
this effect.  A lower cut-off is already applied due to the intrinsic
scatter in the orientation measurements of near circular images.

A potential worry is that we rely on an {\it ellipticity} distribution
taken from ground-based data limited at $I \sim 23$ taken in $\simeq$
1 arcsec seeing (Colless \& Ellis' unpublished NTT data). Clearly
there could be some degradation of the source ellipticities. Although the
HST MDS data covers a smaller area, a two sample K-S test shows a
85\% probability that the two ellipticity distributions came from the
same parent population.

Finally, there is the question of the accuracy of the scale length
distribution used. A similar concern arises in the observations of
giant arcs (Wu \& Hammer 1992, Smail \et 1993).  Unfortunately the small area
and brighter limiting magnitude of the MDS data makes a definitive
statement impossible. The derived scale lengths from the HST
data ($r \sim $0.3-0.7 arcsec at $I=23$) are comparable to those
measured from the cluster observations when allowance has been made
for the effect of seeing. When applied to the fainter $I\sim25$ sample
this technique gives a similar range of scale sizes.  This quantifies
the visual impression that the model faint galaxy images appear very
similar to the real data.

In the presence of a strong intrinsic lensing signal the observational
effects listed above appear to bias only the derived velocity
dispersion of the cluster while returning a good estimate of the input
core radius.  The measured bias then amounts to a velocity dispersion
offset of $-$300-400 km sec$^{-1}$ between the input and output cluster
parameters and no offset in the derived core radius.  For weak lenses
it becomes increasingly difficult to distinguish the offset because of
the weak differentiation between the allowed model parameters.

\section{Results and Discussion}

In discussing the results, we will first present the basic evidence for
gravitational lensing in our sample,
and examine its dependence on the colour of the faint field population.
Here we will, essentially, follow the original method used by Tyson
\et (1990) and utilise the orientations of faint galaxies perpendicular
to the cluster radius vector. Some constraints on the
mean redshift of the $I\leq 25$ sample is possible from this analysis,
particularly if we introduce mass estimates for 1455+22 from X-ray
studies. We then apply the more rigorous maximum likelihood analysis to
examine the datasets for all 3 clusters in the context of the 3 model
$N(z)$ discussed in $\S$4. The strength with which we can rule out
various redshift distributions depends on how much freedom we are
willing to assign to the cluster parameters. The reader is again
referred to the companion article (Paper II) for further information.

\subsection{Orientation Histograms}

Tyson \et (1990) introduced the simple test of measuring the fraction of
galaxies aligned tangentially to the cluster radius vector. We start
by analysing such histograms for each of the clusters in turn. Making
the gross simplification that our clusters are identical objects and ignoring
amplification bias, we directly infer the fraction of galaxies behind each
cluster and hence $N(z)$.

We select galaxies in elliptical annuli aligned with those defined by
the cluster members (see Paper II) in order to remove the effect of the
lens ellipticity on the orientation histogram. The centres used
are the optically defined centres of the clusters.  Paper II demonstrates
that these centres are consistent with those defined by both the mass
and X-ray gas distributions in our cluster sample. Figure~9 shows the
resulting orientation histograms for the $I\leq25$ field samples for each
of the three clusters. Both of the lower redshift clusters show an obvious
excess of tangentially-aligned images. In 1455+22 the excess is approximately
190 in a total of 810 objects (23$\pm$2\%) corresponding to a surface
density of 7.0 galaxies arcmin$^{-2}$, whereas for 0016+16 it is $\simeq$80
out of 356 galaxies (21$\pm$2\%) or 4.5 arcmin$^{-2}$. 1603+43 shows no
alignment excess.

Under the assumption of identical clusters, we can compare these
orientation histograms with our three hypothetical $N(z)$ distributions.
Although considerably idealised, this is illustrative in determining
which cluster is the most critical in estimating $N(z)$. Using a K-S test to
compare the real and analytical orientation distributions, we rule out the
Shallow distribution at the 99.7\% level using the combined result from all
three clusters. As expected, 0016+16 provides by far the strongest rejection
since, according to the Shallow $N(z)$, only 2\% of the $I\leq$25
population should be beyond $z$=0.55. Were 1603+43 to be as massive
and concentrated as the other clusters, the data would also rule out
the Deep $N(z)$.

The signal in the two lower redshift clusters is sufficiently strong
that we can sub-divide the sample to check Tyson {\it et al.}'s claim that
the the excess is predominantly seen in the bluest or faintest galaxies.
We note, however, that by virtue of using the EEV CCD we only have a \VI\
baseline compared to Tyson {\it et al.}'s  \BR. We chose to split the samples
at the mean sample colour of \VI\ = 1.5 and $I_{iso} = 23$ -- where the
colour distribution rapidly begins to shift to the blue (Figure~5(d)).

The orientation histograms for the four sub-classes are shown in
Figure~10(a) for 1455+22 and Figure~10(b) for 0016+16. Clearly all four
sub-classes for both clusters show similar alignments and so we cannot
improve the signal contrast by applying photometric selections. The
photometric distribution of galaxies in the aligned bins ($\theta > 60$
degrees) are completely consistent with those in the unaligned bins. Indeed,
for 1455+22, the two most obvious arclets have colours which fall either
side of that of the cluster members, and the radial distribution of the
aligned component shows no variation of colour with radius.

We can place a strict limit on the maximum possible colour difference
for the {\it excess} aligned population if we assume that they are drawn from
a population with a colour distribution similar in shape to that
observed in the unaligned bins but shifted to the blue. The upper
limits then refer to the maximum colour shift allowed. For 1455+22 the
90\% confidence limit is $\Delta$(\VI) $= -0.2$ and for 0016+16 it is also
$\Delta$(\VI) $= -0.2$. In other words, the aligned component cannot be
drawn from a population which is much bluer in \VI\ than the main
population. Alternatively, the strong shift to the blue seen in the
field colours beyond $I=23$ is not primarily due to the existence of
a distant blue $z\gs1$ galaxy population. This constraint argues against
generic Deep models containing a primordial population of distant
star forming galaxies unless they exhibit a wide spread in \VI\ colours.

\subsection{Constraints from Current Spectroscopic Surveys}

The relatively strong alignment signal in the samples brighter than $I$=23
prompts us to consider a `boot-strap' method for determining the redshift of
the
faint galaxy population. If we can measure the lens parameters for a bright
sample for which the field redshift distribution is already secure from
conventional spectroscopy, we can then apply the lens model to derive the
median redshift of the $I\leq25$ sample.

At the current time, the deepest $I$-selected field survey
are those of Lilly (1993) and Tresse \et (1993) limited at $I\le22$. To
undertake this analysis, we selected galaxies in 1455+22 and 0016+16
satisfying $I \in [20,22]$ and $\epsilon_{opt} \geq 0.05$. To test the
method, we chose two input redshift distributions: the one observed by
Lilly (hypothesis: $H_0$) and the Shallow distribution (hypothesis:
$H_1$). We applied a combined maximum likelihood estimator on the
radial distribution and orientations and, as expected, 0016+16 is the
better discriminator. Taking both clusters, the Shallow redshift
distribution was rejected at the 95\% level to this apparent magnitude
limit. The maximum likelihood parameters for the 1455+22 lens were
$\sigma_{cl}$=1300 km sec$^{-1}$, $r_c$=400 kpc with $\log \hat p_0 =
-0.1$. For 0016+16, $\sigma_{cl}$=1800 km sec$^{-1}$ and $r_c$=70 kpc
with $\log \hat p_0 = -1.9$.  As these estimates are derived from analysis
of a sample of bright galaxies they are less affected by the systematic
biases detailed in $\S$4.3.

The uncertain cluster parameters inferred from the small samples that
overlap in apparent magnitude with the current spectroscopic redshift
surveys indicate that this `boot strap' technique is probably too
hazardous a method for estimating $N(z)$ to $I\leq25$. We will see
later, however, that the derived cluster parameters are not that
erroneous. The prospect of deeper surveys (and thus higher galaxy
surface densities) from 10-m class telescopes will  make the `boot
strap' method a viable approach in future.

\subsection{Maximum Likelihood Analyses}

We now apply the likelihood analysis described in $\S$4.3 to the
datasets constructed from the three clusters. We use the complete $I\in
[20,25]$ samples restricting the FOCAS ellipticities to $\epsilon \in
[0.1,0.8]$.  The likelihood test compares the observed
$r,\theta,\epsilon$ distributions with those from 10 combined
realisations of each lens model calculated for each of the three model
$N(z)$. The parameter space searched is:  $\sigma_{cl} \in [400,2000]$
km sec$^{-1}$ and $r_c \in [0,2]$ arcmin.  The upper limit on the core
radius translates into metric radii of 0.6, 0.9 and 1.0 Mpc for the 3
clusters. The rather small lower limit for the cluster velocity
dispersion protects against the probability peak moving outside our
searched parameter range due to the degradation illustrated in
Figure~8.

For 1455+22 the likelihood distributions (Figure~11(a)) have very similar
shapes to those seen in the simulations presented earlier (Figure~8). The
maximum probabilities for the three redshift distributions are:
$\hat p_{Shallow} = -1.1$, $\hat p_{NE} = -0.7$ and $\hat p_{Deep} = -0.8$.
Thus, the no evolution $N(z)$ is marginally preferred. The lens
parameters for this fit are: $\sigma_{cl} =$ 630$\pm$150 km sec$^{-1}$ and
$r_c =$ 210$\pm$100 kpc.  The best fit velocity dispersion is strictly a
lower limit to the actual value due to the systematic effects
illustrated in the simulations above.  The errors quoted are 90\%
confidence limits taking the errors in $\sigma_{cl}$ and $r_c$ to be
orthogonal.

As before, 0016+16 (Figure~11(b)) provides the most
significant constraint. Here we obtain $\hat p_{Shallow} = -6.2$, $\hat
p_{NE} = -2.8$ and $\hat p_{Deep} = -3.3$ ruling out the Shallow model
and again preferring the no evolution $N(z)$. The lens parameters for
the N.E.\ fit are:  $\sigma_{cl} =$ 860$\pm$250 km sec$^{-1}$ and $r_c =$
210$\pm$250 kpc, where again $\sigma_{cl}$ represents a lower limit to the
actual value.

For 1603+43, the three models have maximum probabilities of:  $\hat
p_{Shallow} = -1.2$, $\hat p_{NE} = -1.3$ and $\hat p_{Deep} = -1.4$.
Given the uncertain cluster mass a very wide parameter space was
searched. For the N.E.\ model a large parameter space is compatible
with the observations (Figure~11(c)).  For both the N.E.\
and Deep models we obtain $\sigma_{cl}
\simeq$ 400 km sec$^{-1}$  and $r_c \simeq$ 900 kpc for the maximum
likelihood solutions.  However, it is
apparent from Figure~11(c) that if a lower limit is placed on the
velocity dispersion of this cluster or an upper limit on its core
radius, we would strongly reject the Deep redshift distribution.

Combining all three clusters we can reject the Shallow redshift distribution at
the 99.98\% level (3.8$\sigma$) but without further assumptions about the
cluster properties we can only claim a marginal preference for the N.E.\ model
over the Deep distribution at the level of 23\% (1.2$\sigma$).

We can refine the derived cluster parameters for the lower redshift
clusters from simulated frames ($\S$4.3.2). The relatively strong
signal in these systems gives a well-determined transformation. From
the likelihood fits for the no evolution $N(z)$ we already have lower
bounds of $\sigma_{cl} \sim 600$ km sec$^{-1}$ and 850 km sec$^{-1}$ for
1455+22 and 0016+16 respectively.  We simulated images for a family of
clusters with different lens parameters ($\sigma_{cl}$, $r_c$) at the
two cluster redshifts ($z=0.26$, $z=0.55$) using the N.E.\ redshift
distribution. The likelihood analysis was then run on FOCAS catalogues
created from each of these images and the input lens parameters of the
simulation whose measured parameters are closest to those observed then
provide the final corrected estimates of $\sigma_{cl}$ and $r_c$ for
the two clusters.

For 1455+22 the simulation whose derived lens parameters lay closest to the
observations had $\sigma_{cl} =$ 1000 km sec$^{-1}$ and a core radius of $r_c
=$ 300 kpc as input. The systematic offset of $-$300-400 km sec$^{-1}$ in the
dispersion is similar to that shown in Figure~8. However, as before, the best
fitting core radius appears to be representative of the intrinsic core size.
Singular potentials do not appear to be able to recreate the observed lens
parameters for a reasonable choice of input parameters. The cluster velocity
dispersion is below that inferred from the X-ray data but above the
spectroscopic value.

In 0016+16 the closest matched simulation has an input dispersion of
$\sigma_{cl} =$
1200 km sec$^{-1}$ and a $r_c =$ 300 kpc core. Again, singular models fail to
reproduce the observed lens parameter values. The systematic offset in
$\sigma_{cl}$
is close to +$400$ km sec$^{-1}$ and $r_c$ is in rough agreement with those
previously calculated for both the mass and galaxy distributions.  The
corrected velocity dispersion is gratifyingly close to  the spectroscopic value
measured by Dressler \& Gunn (1992).

\subsection{Redshift Distributions}

Our primary conclusion is the rejection of a redshift distribution to
$I\leq$25 significantly {\it shallower} than the no evolution prediction.
Such a model would be expected if the ultra-faint counts were dominated by
a population of low redshift ($z<$0.5) dwarfs. The detailed dynamical and/or
X-ray data on our two lower redshift clusters, endorsed by the lensing maps
presented in Paper II, makes this a very robust conclusion. The lensing
signal seen in 0016+16 alone is consistent with the bulk (70-80\%) of the
field galaxies to $I\leq25$ lying beyond the cluster ($z\gs0.6$).

To constrain redshift distributions {\it deeper} than the no evolution
form, we have to consider our most distant cluster, 1603+43. While
less information is available about this cluster than for the two lower
redshift systems, precision data is not so important as {\it any} lensing
signal would indicate the presence of a distant population ($z\gs2$) to
$I\leq25$. The absence of any signal suggests that a substantial ($>$20\%)
tail beyond $z\simeq$2 (such as implied in the White \& Frenk model)
is unlikely. To formally reject a high redshift component at any level,
however, a lower limit is needed on the velocity dispersion of 1603+43.
At present, this can only be derived from X-ray observations, which,
interpreted in terms of low redshift correlations, predict $\sigma_{cl}>$
800 km sec$^{-1}$ for an assumed $r_c<$250 kpc. Adopting the nominal
offset to the model velocity dispersions discussed in $\S$4.3.2, these
parameters for 1603+43 would allow us to reject the 20\% $z>$2 tail at
a confidence limit of $>$95\%.

Support for our conclusion of an absence of a high-$z$ component to the
$I\leq25$ ($B\ls27$) redshift distribution is provided by the  redshift
distribution recently derived by Kneib \et (1993) for $\sim$40 arclets
seen through the rich cluster Abell 370.  The median redshift they
obtain for their sample is $z \sim 0.9$, very close to that of the no
evolution model (Figure~7).  Only 15\% of their `best' sample have $z
\geq 1.5$.

We can compare the derived line of sight velocity dispersions for the
two lower redshift clusters with the predicted distribution from the
CDM simulations of Frenk \et (1990) rescaled to $b=1$ in line with the
COBE observations.  The high inferred velocity dispersions of the
clusters, especially for 0016+16 (in both $N(z)$ cases) given its
redshift, are at the extremes of the standard cold dark matter
predictions.  This is particularly interesting as the derived
dispersions are free of the projection effects commonly invoked to
force agreement between the predicted cluster velocity dispersion
distribution and that observed locally with spectroscopic samples.
This discussion is extended in Paper II where we compare the 2-D
distributions of the baryonic and total mass in our clusters.  However,
we conclude here that selection of clusters according to X-ray
luminosity does provide massive systems.  Although, the absence of
strongly lensed arcs in these two high dispersion clusters implies that
X-ray selection does not necessarily guarantee a potential compact
enough to produce a giant arc.

Although our results rely to some extent on the nature of the clusters
used as lenses, we demonstrate in Paper II that the mass distribution assumed
for the two lower redshift systems can be directly checked from the lensing
data, with satisfactory conclusions. The prospects with 8-10 metre telescopes
for improving our understanding of the dynamical state of 1603+43
via $\simeq$30-50 redshifts (such as are available for 0016+16) and
for enlarging the overlap between lensing-based and spectroscopically-based
field redshift distributions ($\S$5.2) are excellent. We can thus look
forward to tighter constraints on the statistical distances to faint galaxies
in the near future.

\section{Conclusions}

We have developed new and powerful lensing techniques that are able,
with some limitations, to simultaneously constrain the statistical
distances $N(z)$ to the field population at limits well beyond reach
of current spectrographs {\it and} the distribution of dark matter
in a non-parametric manner. The latter conclusions are discussed in more
detail in the companion article, Paper II. We describe detailed tests
of these techniques which, we believe, ensure that systematic biases
inherent in our observational datasets are well-understood.

Our conclusions at this stage can be summarised as follows:

\noindent{$\bullet$} The strongest constraint we can provide on the redshift
distribution of a sample of $I\leq25$ field galaxies is the absence of a
significant population of faint low-$z$ dwarfs such as might be expected if
either the faint end of the local galaxy luminosity function is seriously
underestimated (c.f.\ Koo \et 1993) or if there is strong evolution
of the faint end slope at low redshifts.
We reject a model $N(z)$ with 98\% of the
$I$=25 population at $z \leq 0.55$ at the 99.98\% level.
The very deep surface brightness limit of our imaging data
makes this a particularly effective constraint.

\noindent{$\bullet$ Our constraints on models with large fractions of
high redshift galaxies are weaker due to the lack of detailed
information about the dynamical state of our most distant cluster,
1603+43.  Further study of this system and the properties of other
high $z$ clusters is necessary to make progress. At present we
are unable to formally reject models with as high as 20\% of the
$I\leq25$ population at $z\sim$2.  However, if as we surmise
from its X-ray properties 1603+43 is compact and massive we can reject
the Deep model at better than a 95\% confidence level.

\noindent{$\bullet$} By incorporating external constraints on
the likely range of parameters for the three clusters, our
prefered redshift distribution is therefore the no evolution model. We might
understand how such a model occurs, notwithstanding the high number counts,
if the galaxy luminosity function evolves in shape according to
the empirical form described by  Broadhurst \et (1988, see their Figure 4).
A possible astrophysical model of such evolution incorporating mergers
of fragments whose star-formation rate is slowly declining has been described
by Broadhurst \et (1992).

\noindent{$\bullet$} The lensing strength of our two lower redshift
clusters provides a direct measure of the probable velocity dispersion
($\sigma_{cl}$) and core radii ($r_c$).  However, the estimates,
particularly $\sigma_{cl}$, are affected by systematic biases.  Using
detailed simulations, we correct the biases and for 1455+22 obtain
$\sigma_{cl}$=1000$\pm$200 km sec$^{-1}$ (compared to the spectroscopic
value of $\sim$ 700 km sec$^{-1}$) and, for 0016+16,
$\sigma_{cl}$=1200$\pm$300 km sec$^{-1}$ (compared to the spectroscopic
value of 1300 km sec$^{-1}$). We attempt to derive constraints on the
cluster parameters using the lensing signal to $I \leq 22$ where the
field $N(z)$ is already well-understood from redshift surveys. Whilst
the values derived are consistent with those above, no improved
estimates are obtained due to the low surface density of galaxies
available at the current spectroscopic limit.
We conclude that both clusters are on the extreme
tail of the predicted distribution of cluster dispersions in standard
CDM (Frenk \et 1990).

\section*{ACKNOWLEDGEMENTS}

The above project was initiated with inspiration derived from
the enthusiasm of Bernard Fort, Yannick Mellier and Genevieve Soucail
at Toulouse Observatory and the pioneering work of Tony Tyson
and colleagues. We acknowledge useful discussions with Gary Bernstein,
Carlos Frenk, Nick Kaiser, Chris Kochanek and Jordi Miralda-Escud\'e.
The support of the La Palma staff and especially Dave Carter is also
gratefully acknowledged. This work was supported by the SERC.

%\begin{listoffigures}
\noindent{\bf FIGURES}
\smallskip

\noindent{{\bf Figure 1:}
A 1$\times$1 arcmin test area in the 1455+22 frame. This is a 20.8
ksec $I$  exposure with 40 objects detected above $I=25$.  Those
objects in the faintest analysis sample ($I_{iso}\in 24-25$) are
marked.}

\noindent{{\bf Figure 2:}
Composite $V+I$ images of the clusters: (a) 1455+22 ($z$=0.26), (b)
0016+16 ($z$=0.55) and (c) 1603+43 ($z$=0.89).  The total exposure
times are 33.0, 36.5 and 45.8 ksec respectively and the scales are in
arcseconds with east to the left and north at top.}

\noindent{{\bf Figure 3:}
The \VI\ colour-magnitude diagrams for the clusters (a) 1455+22, (b)
0016+16 and (c) 1603+43.  Galaxies with colours similar to an E/S0
at the cluster redshift are shown ($\bullet$), as well as the 80\%
and 50\% completeness limits for the various catalogues, calculated
from simulations. Also shown are the photometric errors as a function
of magnitude for the entire samples, the line marked is the chosen
magnitude limit for the total sample $I=25$.}

\noindent{{\bf Figure 4:}
Differential number counts in $I$ of (a) 1455+22, (b)
0016+16 and (c) 1603+43 after removal of the cluster members
($\bullet$) compared to the Lilly \et (1991) field counts ($\circ$).
The deficit of field galaxies in the more distant clusters at faint
magnitudes arises from the cluster selection criteria.

\noindent{{\bf Figure 5:}
(a)-(c) \VI\ Colour distributions for the clusters (filled) and field
galaxies (open) brighter than $I=25$ in the three clusters.  (d) The
\VI\ colour distribution of the combined field samples from the three
clusters, split into the various magnitude slices.  The bars at the top
show the range of colours covered by  the non-evolved morphological
types as a function of redshift. The galaxy colours generally start at
the left side for $z=0$ and move right until $z\sim 0.5$ at which point
they start becoming bluer again.}

\noindent{{\bf Figure 6:}
The two panels show the comparison of the raw (FOCAS) and  optimally
weighted  ellipiticites for the 1455+22 test area.  The top panel
illustrates the systematic offset (dotted line) introduced in the ellipticity
measurement when using the optimal weighting scheme.  The lower panel
compares the  ellipticities of objects measured on the two independent
frames.  It is apparent that the optimal weighting reduces the scatter
in individual ellipticity measurements for $I_{iso} \in [24,25]$ objects.}

\noindent{{\bf Figure 7:}
(a) The various normalised redshift distributions used in the
analysis.  The dashed curves marked WF25 and WF27 are the $B = 25$ and
$B = 27$ distributions from White \& Frenk 1991.  The curve shown as
dotted is a Bruzual $B = 27$ cumulative $N(z)$.  The remaining solid
curves show the no evolution differential $N(z)$ centred on the $I$
magnitude marked -- these were calculated for observations in $R$ band
and then converted using a fixed colour term. (b) The run of median
redshift with $I$ magnitude for the three hypotheses.}

\noindent{{\bf Figure 8:}
The top panels show the  likelihood contours of a test of an analytic
simulation while the bottom panels show the contours for a numerical
simulation with the same lens parameters. The probability contours are
logarithmic and spaced every factor of 10 -- starting at 10$^{-10}$
below the peak probability. Details of the  simulations are given in
the text, the input redshift distribution was the no evolution model
and the  lens parameters are marked ($\bullet$).  the core radius
($r_c$) is in kpc and the velocity dispersion ($\sigma_{cl}$) is in km
sec$^{-1}$.}

\noindent{{\bf Figure 9:}
Orientation histograms for 1455+22, 0016+16 and 1603+43. The
histograms were constructed using all objects in the field samples with
$I \in [20,25]$ and \VI $\in [-1,2.5]$ with optimally weighted
ellipticities above a cutoff of $\epsilon \geq 0.05$.  The orientations
and ellipticities used for the annular bins were those quoted in the
text and the optically determined lens centres were used.}

\noindent{{\bf Figure 10:}
(a) The orientation histograms for 1455+22 separated in terms of colour
and magnitude.  The aligned excess has a similar strength in all four
samples showing that it has no strong colour or magnitude dependency.
(b) As above for the 0016+16 sample.  As in
(a) all four samples contain an excess of tangentially
aligned objects confirming that this arises from a population of
objects with similar characteristics to the bulk of the faint field
population.}

\noindent{{\bf Figure 11:}
The likelihood distributions for the cluster lensing models of
(a) 1455+22, (b) 0016+16 and (c) 1603+43.
Each plot is for one of the model redshift distributions.
The units for the axes are km sec$^{-1}$ and kpc.  The contours are for the
smoothed probability distributions and are spaced every factor of 10
from the peak.}

%\end{listoffigures}

\end{document}